# Bending The Heisenberg Uncertainty Principle

Anwar Mohiuddin[a], Abhijeet K. Jha[b] and Prasanta K. Panigrahi[a]

[a]IISER − Kolkata, Mohanpur Campus, Nadia, West Bengal − 741252
[b]IISER − Pune, Sutarwadi Road, Pashan, Pune, Maharastra − 411021

The celebrated Heisenberg Uncertainty Principle Δx.Δp≥ ħ/2 can allow measurement accuracies less than Δx or Δp. Classical analog of this is known as sub-Fourier sensitivity. We illustrate this phenomenon in a step by step process using the example of compass state, as suggested by Zurek.

A number of canonically conjugate variables appear in mechanics, like co-ordinate-momentum and time-energy. The fact that, they are related through Fourier transform, restricts their measurement accuracies. For example, it is well-known from the theory of Fourier transform that,

$$\Delta x.\Delta k \sim 1,$$

where the Fourier transform of a function $F(x)$ of the co-ordinate variable is related to its Fourier counterpart $\tilde{F}(k)$ in the form,

$$F(x) = \int \frac{d^3k}{(2\pi)^3} e^{ikx}\, \tilde{F}(k).$$

In quantum mechanics, the above uncertainty product leads to the Heisenberg uncertainty relation,

$$\Delta x.\Delta p \geq \frac{\hbar}{2},$$

where $p = \hbar k$.

For a Gaussian state of the type,

$$\psi(x) = \left(\frac{m\omega}{2\hbar}\right)^{\frac{1}{4}} e^{-\frac{m\omega\, x^2}{2\hbar}},$$

familiar from the harmonic oscillator problem, the uncertainty relation leads to an equality,

$$\Delta x.\Delta p = \frac{\hbar}{2}.$$

Explicit calculation yields,

$$\Delta x = \sqrt{< x^2 > - < x >^2} \equiv \sqrt{< x^2 >} = \sqrt{\frac{\hbar}{2m\omega}},$$

and,

$$\Delta p = \sqrt{< p^2 > - < p >^2} \equiv \sqrt{< p^2 >} = \sqrt{\frac{\hbar m\omega}{2}}.$$

Here, $< x^2 > = \int_{-\infty}^{\infty} \psi^*(x) x^2 \psi(x) dx$ and $< p^2 > = \int_{-\infty}^{\infty} \psi^*(x)(-\hbar^2 \frac{\partial^2}{\partial x^2})\psi(x) dx$.

By use of more general Gaussian states, like squeezed states, one can reduce one of the uncertainties:

$$\Delta x \to \frac{\Delta x}{\lambda}, \text{ and } \Delta p \to \lambda \Delta p,$$

maintaining $\Delta x . \Delta p = \frac{\hbar}{2}$. It is then natural to ask, if such states exist for which it is possible to measure variation in $x$ (or $p$), which is less than $\Delta x$ (or $\Delta p$). In Fourier transform, this is known as sub-Fourier sensitivity and has been experimentally demonstrated recently, through appropriate combination of laser beams [1]. In the quantum domain, it was demonstrated by Zurek [2], that the above can be achieved through special states like cat and compass states. These states are superposition of familiar Gaussian states and hence the reason behind this sensitivity can be appreciated without tedious effort. The following problem illustrates this, in a step by step process.

Q1) Show that the displaced Gaussian function $e^{-(x-\alpha)^2/2} \equiv < x \mid \alpha >$ is an Eigen state of $\mathrm{a} = \left(x + \frac{\partial}{\partial x}\right)$, with Eigen value$\alpha$.

Proof: Since a $< x \mid \alpha > = \left(x + \frac{\partial}{\partial x}\right) e^{-(x-\alpha)^2/2}$

$= x e^{-(x-\alpha)^2/2} + \frac{\partial}{\partial x} e^{-(x-\alpha)^2/2}$

$= \alpha e^{-(x-\alpha)^2/2}$, the displaced Gaussian function.

$e^{-(x-\alpha)^2/2}$ is an Eigen state of a. It is worth noting that $< x \mid \alpha >$ is known as the coherent state in literature, which describes laser. A discerning reader will recognize that modulo constant factors, a is the annihilation operator of the harmonic oscillator problem.

Q2) Given that$\psi = N(e^{-\frac{(x-\alpha)^2}{2}} + e^{-\frac{(x+\alpha)^2}{2}})$, find out the normalization constant N from the square integrability condition: $\int_{-\infty}^{\infty} \psi^* \psi dx = 1$.

Hint: one can take $\alpha$ to be real and use the formula $\int_{-\infty}^{\infty} e^{-ax^2} dx = \sqrt{\frac{\pi}{a}}$ .

Solution: assuming N and $\alpha$ to be real;

$$\int_{-\infty}^{\infty} \psi^* \psi \, dx = \int_{-\infty}^{\infty} \left( N \left( e^{-\frac{(x-\alpha)^2}{2}} + e^{-\frac{(x+\alpha)^2}{2}} \right) \right)^* \left( N \left( e^{-\frac{(x-\alpha)^2}{2}} + e^{-\frac{(x+\alpha)^2}{2}} \right) \right) dx = 1$$

$$= N^2 \int_{-\infty}^{\infty} \left( e^{(x-\alpha)^2} + e^{(x+\alpha)^2} + 2e^{-(x^2+\alpha^2)} \right) dx = 1$$

Substituting $x - \alpha = y$ in the first expression and carrying out similar manipulations in the last two expressions, the above integrals can be straightforwardly evaluated and one obtains

$$2\sqrt{\pi} N^2 \left(1 + e^{-\alpha^2}\right) = 1 \text{, yielding } N = \left(\frac{1}{\pi}\right)^{\frac{1}{4}} \left(\frac{1}{2\left(1+e^{-\alpha^2}\right)}\right)^{\frac{1}{2}}.$$

Q3) Given that $\phi = \left( e^{-\frac{(x-\alpha)^2}{4\sigma^2}} + e^{-\frac{(x+\alpha)^2}{4\sigma^2}} \right) e^{-ikx}$ and

$\phi_\delta = \left( e^{-\frac{(x-\alpha)^2}{4\sigma^2}} + e^{-\frac{(x+\alpha)^2}{4\sigma^2}} \right) e^{-ikx} e^{i\delta x}$, Calculate the overlap integral, $I = \int_{-\infty}^{\infty} \phi_\delta^* \phi \, dx$,and find out the points it vanishes. Give physical interpretation for this phenomenon.

Solution: Taking $\alpha$ to be real, for simplicity, one finds

$$I = \int_{-\infty}^{\infty} \left( \left( e^{-\frac{(x-\alpha)^2}{4\sigma^2}} + e^{-\frac{(x+\alpha)^2}{4\sigma^2}} \right) e^{-ikx} e^{i\delta x} \right)^* \left( \left( e^{-\frac{(x-\alpha)^2}{4\sigma^2}} + e^{-\frac{(x+\alpha)^2}{4\sigma^2}} \right) e^{-ikx} \right) dx$$

$$= \int_{-\infty}^{\infty} \left( e^{-((x^2+\alpha^2-2x\alpha+2\sigma^2 i\delta x)/2\sigma^2)} + e^{-((x^2+\alpha^2+2x\alpha+2\sigma^2 i\delta x)/2\sigma^2)} + 2e^{-((x^2+\alpha^2+2\sigma^2 i\delta x)/2\sigma^2)} \right) dx.$$

We now consider each term individually:

$$1^{\text{st}} \text{ term} = \int_{-\infty}^{\infty} e^{-((x^2+\alpha^2-2x(\alpha-\sigma^2 i\delta)+(\alpha-\sigma^2 i\delta)^2-(\alpha-\sigma^2 i\delta)^2)/2\sigma^2)} \, dx$$

Redefining the variable as $x - \alpha + \sigma^2 i\delta / \sqrt{2}\,\sigma = z$, and using the above mentioned result we obtain

$$\sqrt{2\pi}\sigma e^{-\sigma^2\delta^2/2} e^{i\alpha\delta}.$$

Similarly we get the second and third terms as $\sqrt{2\pi}\sigma e^{-\sigma^2\delta^2/2}e^{-i\alpha\delta}$ and $2\sqrt{2\pi}\sigma e^{-\sigma^2\delta^2/2}e^{-\frac{\alpha^2}{2\sigma^2}}$ respectively.

We note that the third term is completely real as compared to the first two terms in the integral.

Adding the results leads to,

$$I = \sqrt{2\pi}\sigma e^{-\frac{\sigma^2\delta^2}{2}}\left(e^{-i\delta x} + e^{i\delta x} + 2e^{-\frac{\alpha^2}{2\sigma^2}}\right).$$

We also note that the first two terms in the above result lead to an oscillatory factorcos(δα), where as the third term led to Gaussian factor $e^{-\alpha^2/2\sigma^2}$. It can be easily seen that the above expression vanishes when $\cos(\alpha\delta) = -e^{-\alpha^2/2\sigma^2}$. At these points $\phi$ and $\phi_\delta$ are orthogonal and hence can be distinguished from each other. We would like to emphasize that only orthogonal states can be perfectly distinguished from each other. This indicates that through the above interferometric arrangement a shift $\delta$ in $k$ can be determined. This was first suggested by Zurek[2] and has been experimentally verified by Praxmeyer et al. [1], in a laser set up.

Q4) for the normalized wave function$\psi = \left(\frac{1}{2\sigma\sqrt{2\pi}\left(1+e^{-\frac{\alpha^2}{2\sigma^2}}\right)}\right)^{\frac{1}{2}}\left(e^{-\frac{(x-\alpha)^2}{4\sigma^2}} + e^{-\frac{(x+\alpha)^2}{4\sigma^2}}\right)e^{-ikx}$,

calculate the uncertainties $\Delta x$ and $\Delta k$ and check that $\delta$ is smaller than $\Delta k$! Interpret your result keeping in mind the Heisenberg's uncertainty principle.

Solution: We note that $\Delta x = \sqrt{< x^2 > - < x >^2}$

$$< x^2 > = \int_{-\infty}^{\infty} \psi^* x^2 \psi dx$$

$$= \frac{1}{2\sigma\sqrt{2\pi}\left(1+e^{-\frac{\alpha^2}{2\sigma^2}}\right)}\int_{-\infty}^{\infty}\left(\left(e^{-\frac{(x-\alpha)^2}{4\sigma^2}} + e^{-\frac{(x+\alpha)^2}{4\sigma^2}}\right)e^{-ikx}\right)^* x^2 \left(e^{-\frac{(x-\alpha)^2}{4\sigma^2}} + e^{-\frac{(x+\alpha)^2}{4\sigma^2}}\right)e^{-ikx}\, dx$$

We now consider each term obtained by simplification of the above expression separately:

1st term $= \int_{-\infty}^{\infty} e^{-\frac{(x-\alpha)^2}{2\sigma^2}} x^2 dx = \sigma\sqrt{2\pi}\,(\sigma^2 + x^2)$.

2nd term $= \int_{-\infty}^{\infty} e^{-\frac{(x+\alpha)^2}{2\sigma^2}} x^2 dx = \sigma\sqrt{2\pi}\,(\sigma^2 + x^2)$

$3^{\text{rd}}$ term $= \int_{-\infty}^{\infty} 2e^{-(x^2+\alpha^2)/2\sigma^2} x^2 dx = 2\sqrt{2\pi}\,\sigma^3 e^{-\frac{\alpha^2}{2\sigma^2}}$

Adding the results we get $< x^2 > = \sigma^2 + \frac{\alpha^2}{1+e^{-\frac{\alpha^2}{2\sigma^2}}}$.

One finds that $< x > = 0$ for this function since it is symmetric about $\alpha$.

Hence we get $\Delta x = \sqrt{\sigma^2 + \frac{\alpha^2}{1+e^{-\frac{\alpha^2}{2\sigma^2}}}}$.

We now similarly calculate $\Delta p = \hbar \Delta k$.

Here $p$ is the momentum operator, given by $p = \frac{\hbar}{i}\frac{\partial}{\partial x}$.

$$< p > = \int_{-\infty}^{\infty} \psi^* \frac{\hbar}{i}\frac{\partial}{\partial x}\psi dx$$

$$= \frac{1}{2\sigma\sqrt{2\pi}\left(1+e^{-\frac{\alpha^2}{2\sigma^2}}\right)} \int_{-\infty}^{\infty} \left(\left(e^{-\frac{(x-\alpha)^2}{4\sigma^2}} + e^{-\frac{(x+\alpha)^2}{4\sigma^2}}\right) e^{-ikx}\right)^* \frac{\hbar}{i}\frac{\partial}{\partial x}\left(e^{-\frac{(x-\alpha)^2}{4\sigma^2}} + e^{-\frac{(x+\alpha)^2}{4\sigma^2}}\right) e^{-ikx}\, dx$$

We first determine the function obtained after momentum operator operates on $\psi$:

$$= \frac{\hbar}{i}\frac{\partial}{\partial x}\left(e^{-\frac{(x-\alpha)^2}{4\sigma^2}} + e^{-\frac{(x+\alpha)^2}{4\sigma^2}}\right) e^{-ikx} = \frac{\hbar}{i}\left\{e^{-\frac{(x-\alpha)^2}{4\sigma^2}}\left(-ik - \frac{x-\alpha}{2\sigma^2}\right) + e^{-\frac{(x+\alpha)^2}{4\sigma^2}}\left(-ik - \frac{x+\alpha}{2\sigma^2}\right)\right\} e^{-ikx}$$

Hence we get

$$< p > = \frac{\hbar}{i}\frac{1}{2\sigma\sqrt{2\pi}\left(1+e^{-\frac{\alpha^2}{2\sigma^2}}\right)} \int_{-\infty}^{\infty} \left(e^{-\frac{(x-\alpha)^2}{4\sigma^2}} + e^{-\frac{(x+\alpha)^2}{4\sigma^2}}\right)\left\{e^{-\frac{(x-\alpha)^2}{4\sigma^2}}\left(-ik - \frac{x-\alpha}{2\sigma^2}\right) + e^{-\frac{(x+\alpha)^2}{4\sigma^2}}\left(-ik - \frac{x+\alpha}{2\sigma^2}\right)\right\} dx$$

$$= -\hbar k \frac{1}{2\sigma\sqrt{2\pi}\left(1+e^{-\frac{\alpha^2}{2\sigma^2}}\right)}\left(2\sqrt{2\pi}\,\sigma + 2\sqrt{2\pi}\,\sigma\, e^{-\frac{\alpha^2}{2\sigma^2}}\right) = -\hbar k.$$

Now $< p^2 > = \int_{-\infty}^{\infty} \psi^*(-\hbar^2)\frac{\partial^2}{\partial x^2}\psi dx$

Hence $< p^2 > = \left(\hbar^2 k^2 + \frac{\hbar^2}{2\sigma^2} - \frac{\alpha^2\hbar^2 e^{-\frac{\alpha^2}{2\sigma^2}}}{4\sigma^2\left(1+e^{-\frac{\alpha^2}{2\sigma^2}}\right)}\right)$

Hence finally we obtain $\Delta x = \sqrt{\sigma^2 + \frac{\alpha^2}{1+e^{-\frac{\alpha^2}{2\sigma^2}}}}$ and $\Delta p = \sqrt{\frac{\hbar^2}{4\sigma^2} - \frac{\alpha^2\hbar^2}{4\sigma^4(1+e^{-\frac{\alpha^2}{2\sigma^2}})}}$

Yielding the Heisenberg's uncertainty relation $\Delta x \Delta p = \frac{\hbar}{2\sigma^2\left(1+e^{\frac{\alpha^2}{2\sigma^2}}\right)} \sqrt{\sigma^4 \left(1 + e^{\frac{\alpha^2}{2\sigma^2}}\right)^2 - \alpha^4}$

We note that in the experiment of Praxmeyer et. al. [2], the above gaussian functions have been used to describe the laser intensity profiles. The experiment has been carried in time-frequency domain, as compared to co-ordinate-momentum representation used here. Using their experimental numbers in time frequency domain, one finds $\delta = 3.3 THz$ and $\Delta\omega = 4.0 \pm 0.1 THz$..

Hence one clearly sees that $\delta$ can be less than $\Delta\omega$. Although it is counter intutive, Heisenberg's uncertainty relation has not been violated.

References:

[1] L. Praxmeyer, P. Wasylczyk, C. Radzewicz, and K. Wódkiewicz, Phys.Rev.Lett., **98**, 063901(2007).

[2] W. Zurek, Nature (London), **412**, 712 (2001).